\documentclass{webofc}
\makeatletter

\def\blfootnote{\gdef\@thefnmark{}\@footnotetext}

\makeatother
\usepackage[utf8]{inputenc}
\usepackage[varg]{txfonts}   
\usepackage{tikz}
\usepackage{siunitx}
\usepackage{subcaption}
\usepackage{overpic}
\usepackage[capitalise]{cleveref}
\usepackage{tabularx, booktabs}

\usepackage[
    colorinlistoftodos, 
    shadow, 
    figwidth=.75\textwidth, disable]{todonotes}

\begin{document}
\title{Modeling Distributed Computing Infrastructures for HEP Applications\blfootnote{This manuscript has been authored in part by UT-Battelle, LLC, under contract DE-AC05-00OR22725 with the US Department of Energy (DOE). The publisher, by accepting the article for publication, acknowledges that the U.S. Government retains a non-exclusive, paid up, irrevocable, world-wide license to publish or reproduce the published form of the manuscript, or allow others to do so, for U.S. Government purposes. The DOE will provide public access to these results in accordance with the DOE Public Access Plan (http://energy.gov/downloads/doe-public-access-plan).}}
%
%

\author{\firstname{Maximilian} \lastname{Horzela}\inst{1}\fnsep\thanks{\email{maximilian.horzela@kit.edu}} \and
        \firstname{Henri} \lastname{Casanova}\inst{2}\fnsep\thanks{\email{henric@hawaii.edu}} \and
        \firstname{Manuel} \lastname{Giffels}\inst{1}\fnsep\thanks{\email{manuel.giffels@kit.edu}} \and
        \firstname{Artur} \lastname{Gottmann}\inst{1}\fnsep\thanks{\email{artur.gottmann@kit.edu}} \and
        \firstname{Robin} \lastname{Hofsaess}\inst{1}\fnsep\thanks{\email{robin.hofsaess@kit.edu}} \and
        \firstname{Günter} \lastname{Quast}\inst{1}\fnsep\thanks{\email{g.quast@kit.edu}} \and
        \firstname{Simone} \lastname{Rossi~Tisbeni}\inst{3,4}\fnsep\thanks{\email{simone.rossitisbeni@unibo.it}} \and
        \firstname{Achim} \lastname{Streit}\inst{1}\fnsep\thanks{\email{achim.streit@kit.edu}} \and
        \firstname{Frédéric} \lastname{Suter}\inst{5}\fnsep\thanks{\email{suterf@ornl.gov}}
}

\institute{Karlsruhe Institute of Technology, Kaiserstraße 12, 76131 Karlsruhe, Germany 
\and
           University of Hawai'i at Mānoa, 2500 Campus Road, Honolulu, HI 96822-2217, USA 
\and
           Alma Mater Studiorum - Università di Bologna, Via Zamboni, 33, 40126 Bologna BO, Italy
\and
           Istituto Nazionale Di Fisica Nucleare, Viale Berti Pichat, 6/2, 40127 Bologna BO, Italy
\and
           Oak Ridge National Laboratory, 1 Bethel Valley Rd, Oak Ridge, TN 37830, USA
          }

\abstract{%
Predicting the performance of various infrastructure design options in complex federated infrastructures with computing sites distributed over a wide area network that support a plethora of users and workflows, such as the Worldwide LHC Computing Grid (WLCG), is not trivial. Due to the complexity and size of these infrastructures, it is not feasible to deploy experimental test-beds at large scales merely for the purpose of comparing and evaluating alternate designs. An alternative is to study the behaviours of these systems using simulation. This approach has been used successfully in the past to identify efficient and practical infrastructure designs for High Energy Physics (HEP). A prominent example is the Monarc simulation framework, which was used to study the initial structure of the WLCG. New simulation capabilities are needed to simulate large-scale heterogeneous computing systems with complex networks,  data access and caching patterns.
A modern tool to simulate HEP workloads that execute on distributed computing infrastructures based on the SimGrid and WRENCH simulation frameworks is outlined. Studies of its accuracy and scalability are presented using HEP as a case-study. 
Hypothetical adjustments to prevailing computing architectures in HEP are studied providing insights into the dynamics of a part of the WLCG and candidates for improvements.
}
\maketitle

\section{Introduction}
\label{intro}
The physics programs in high energy physics (HEP) at the Large Hadron Collider (LHC) would not be conceivable without the computing infrastructure provided by the Worldwide LHC Computing Grid (WLCG)~\cite{TDR-WLCG,Update-WLCG}.
This complex large-scale distributed computing system is crucial for the processing and storage of the physics data produced by the collaborations at the LHC, e.g., the Compact Muon Solenoid (CMS)~\cite{CMS}.
With the start of the second phase of the LHC program by the end of the decade, the so-called high-luminosity LHC (HL-LHC), an increase by a factor of ten in the data production rate and a proportional increase of the related computing demands are expected.
Such a dramatic increase of the storage and compute requirements will not be met solely by traditional upgrades of the current WLCG resources.
Hence, the available resources will have to be utilized more efficiently.
Identifying changes in the design of an infrastructure as complex, heterogeneous and large as the WLCG, so as to maximize performance and efficiency, is challenging.

To address the above challenge, we propose to study computing infrastructure designs in simulation, which has proved useful in the past for determining the original structure of the WLCG~\cite{MONARC-Simulator,MONARC-HLT}.
Simulation enables the comparison of the execution of HEP applications on several candidate compute and storage infrastructures without the need to build testbeds and without disrupting the execution of workflows on WLCG production systems.
In this paper, we present a novel simulation tool and illustrate its capacity to simulate HEP applications executions on complex large-scale distributed computing systems accurately.

\section{Simulation Tool}
\label{sec:simulator}

We have developed a simulator in C++ to enable the simulation and performance analysis of HEP applications using two state-of-the-art discrete-event open-source simulation frameworks:
SimGrid~\cite{Casanova2014,simgrid-web} and WRENCH~\cite{wrench,wrench-web}. 
The simulation models in both frameworks have been validated~\cite{10.1145/2517448} and make it possible to achieve sensible trade-offs between simulation accuracy and simulation scalability (i.e., speed and memory footprint). 
Hereafter we describe these frameworks, the simulator's implementation, the simulators calibration and validation procedure, and the computational complexity of the simulator.

\subsection{Supporting Simulation Frameworks}
\label{sec:frameworks}

\textbf{SimGrid} provides low-level simulation abstractions, implemented in C++, for the development of simulators of distributed computing systems. Some of these abstractions are used to describe hardware platforms, which comprise network, compute, and storage resources.
Network resources consist of network links, defined by latencies and bandwidths.
Compute resources consist of multicore processing units, defined by numbers of cores and core compute speeds.
Storage resources consist of disks, described by read and write bandwidths.
Routes can be defined between these resources, as vectors of network links, to define the platform's network topology.
Other abstractions are then used to describe simulated processes, or actors that execute on this platform.
Each actor can execute arbitrary C++ code and spawn activities (i.e., communication, computation, or I/O) that utilize hardware resources. A key feature of SimGrid is that it formulates the problem of computing the rate of progress of all simulated activities at a given time as a unified constrained optimization problem.
The constraints involve the activities' rates of progress, scaling factors, and hardware resource capacities.
The objective function corresponds to Max-Min fairness, i.e., maximizing the minimum rate of progress over all activities, where these rates of progress are considered to be real numbers.
This problem must be solved each time an activity starts or completes and SimGrid implements a highly optimized solver that is invoked repeatedly throughout the simulation's execution.
This optimization problem formulation makes it possible to accurately model complex real-world behaviors with low computational complexity~\cite{10.1145/2517448}.

\textbf{WRENCH} builds on SimGrid to ease simulator implementation~\cite{wrench}.
It comes with already implemented abstractions of commonplace components of such systems, which can be used as building blocks for implementing simulators of complex systems, while benefiting from the accurate and scalable simulation models provided by SimGrid.
WRENCH provides implementations for several kinds of services such as compute services (e.g., bare-metal servers, batch-scheduled clusters, HTCondor deployments, cloud infrastructures) and storage services (e.g., simple file servers, proxy file services with caching capabilities, XRootD deployments). 
The user defines a hardware platform and defines which services are deployed on that platform and on which resources.
The user also implements one or more simulation controllers, i.e., processes that interact with the deployed services to execute some application workload of interest.
This interaction is implemented through each service's API.
For instance, WRENCH provides the notion of a job that comprises tasks with arbitrary dependencies, can be submitted to a compute service, and eventually succeeds or fails.

\subsection{Simulator Implementation}
\label{sec:extensions}

At the onset of this project, WRENCH provided all abstractions necessary to simulate the execution of batch computation, where each job simply consists of reading input files, performing computation, and writing output files.
However, such batch jobs make up only a small fraction of the workloads in HEP. Most jobs, instead, are streaming jobs that pipeline input reading, computation, and output writing with a granularity defined, e.g., by the XRootD block size.
To support the simulation of streaming jobs, we implemented in WRENCH the concept of a custom action that allows the user to define a job's behavior, such as streaming, based on an arbitrary C++ function.

Besides streaming jobs, other key behaviors of real-world executions that we wish to simulate are data locality and caching.
Specifically, when attempting to read input data, a job checks a list of cache locations where the data may be available.
If found, then the data is obtained from that location.
Otherwise, the data is obtained from authoritative remote storage locations that hold input data at the onset of the execution. 
In this case, the data is then copied to the caches, which will be able to answer future requests for access to this same data.
When a cache's storage capacity would be exceeded, data is evicted based on some classical eviction policy, such as Least Recently Used (LRU).
We implemented this behavior by using another custom action.
Note that this caching feature is useful in general and, as such, has been implemented in WRENCH as part of the storage service.

WRENCH provides features for monitoring of the simulation. To avoid a too large amount of monitoring data from large-scale executions of our simulator, we implemented our own lightweight simulation monitoring component to track relevant information for our purpose: job identifiers, job execution times, data cache hit rates, and data transfer times.

As explained in the next section, we simulate the execution of multiple concurrent workloads.
A distinct controller is in charge of the execution of each of these workloads.
Each controller interacts with storage services deployed on the platform including XRootD, and with an HTCondor compute service that provides access to and schedules jobs on a set of bare-metal compute services deployed on the platform's compute resources. 

\subsection{Simulator Usage}
\label{sec:simulatortool}

The simulator takes as input: (i)~a description of the compute platform;  (ii)~a specification of the workloads that execute on this platform; and (iii) a specification of the data present in caches at the onset of the execution.
We configure a workload as a set of individual jobs, each executing pipelined input reading and computation based on input file sizes and a data transfer block size, and transferring an output file to a storage.
Each workload is characterized by various parameters, such as the number and type of jobs, the submission time, probability distributions for the number and size of input files, the required CPU cores and memory, a computational volume per byte of input data, and the size of the output file. 

The above input fully characterizes the execution to be simulated, including the scheduling of the jobs and the execution of their underlying activities, which trigger computations, I/O operations, and network communications.
Furthermore, input files are cached when a cache storage location is available in a predefined vicinity of the machine that executes a job, and LRU files are evicted from that cache when full.
When a job completes, its output files are written to the predefined storage locations.
When all jobs have completed, the simulation terminates, and all relevant information to study each individual job is provided.

\subsection{Calibration \& Validation}
\label{sec:calandval}

The simulation models implemented in the simulator are meant to mimic the empirically observed behavior of application executions on real-world platforms. 
So as to capture a range of possible execution scenarios, these models come with free parameters that have to be tuned to best match real-world behaviors.
Picking appropriate values for, or calibrating, these parameters was performed based on ground-truth data.
This data comprises a set of collected execution traces of jobs executions on a controlled, dedicated, and distributed test platform, where the jobs and the platforms have known characteristics.
The calibration process consists in quantitatively comparing these real-world traces to traces generated by the simulator, and adjusting parameter values to achieve best possible match.

\begin{figure}
    \centering
    \begin{overpic}[width=0.49\linewidth]{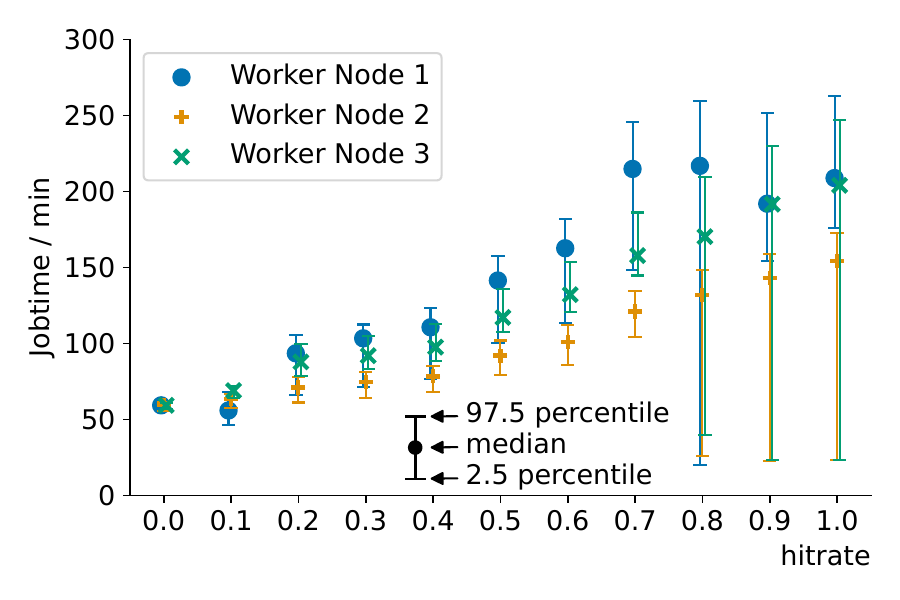}
        \put(50,48){\includegraphics[width=0.09\linewidth]{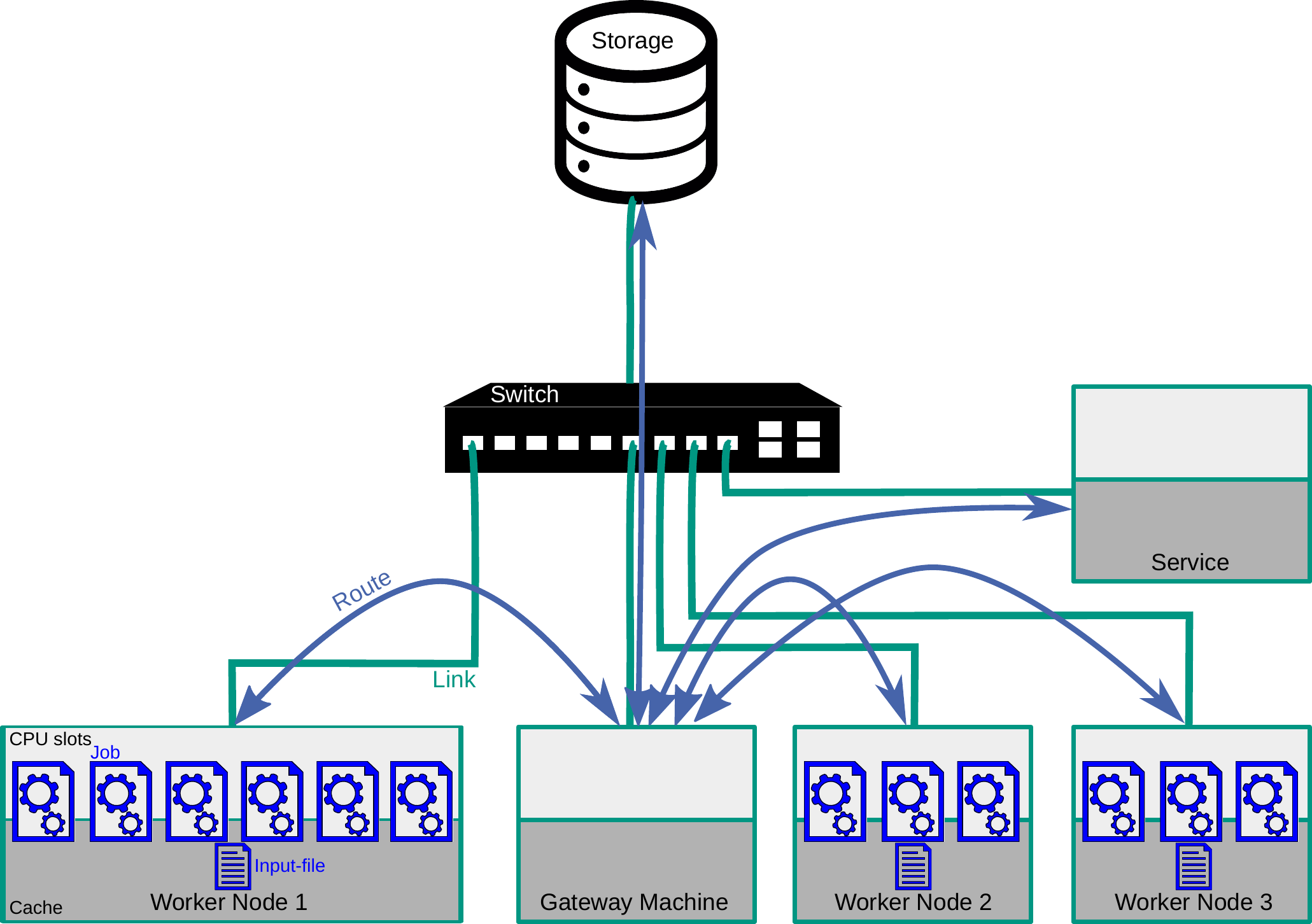}}
    \end{overpic}
    \begin{overpic}[width=0.49\linewidth]{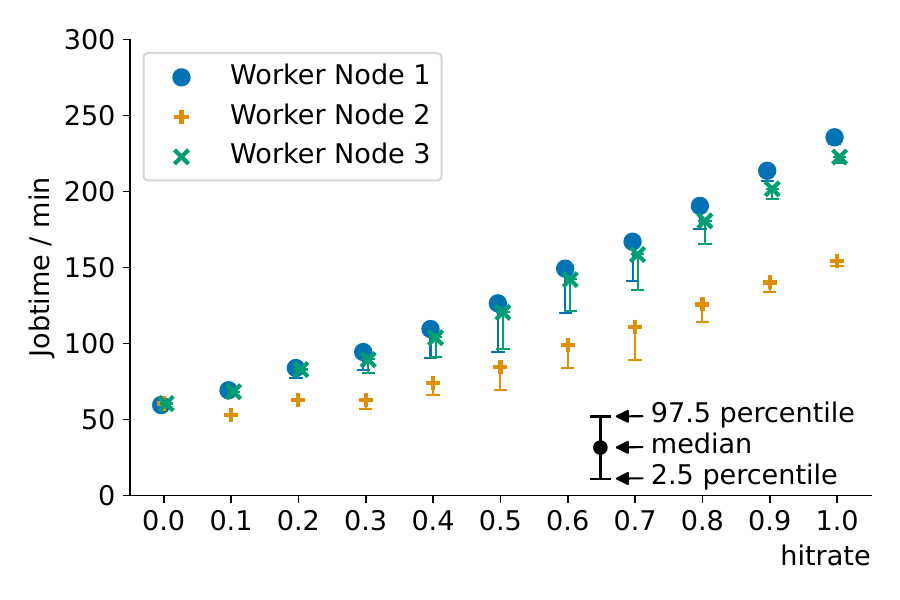}
        \put(50,54){\includegraphics[width=0.05\linewidth]{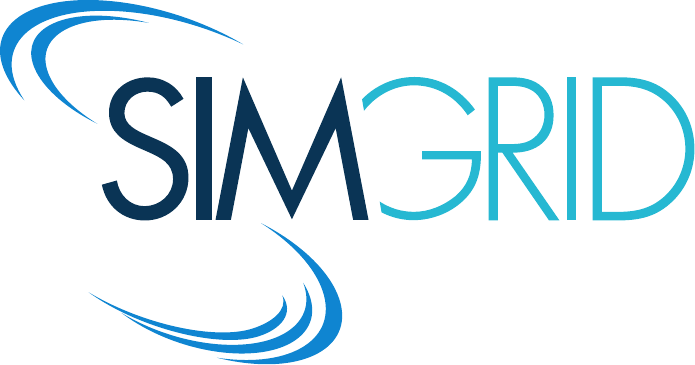}}
        \put(60,50){\includegraphics[width=0.05\linewidth]{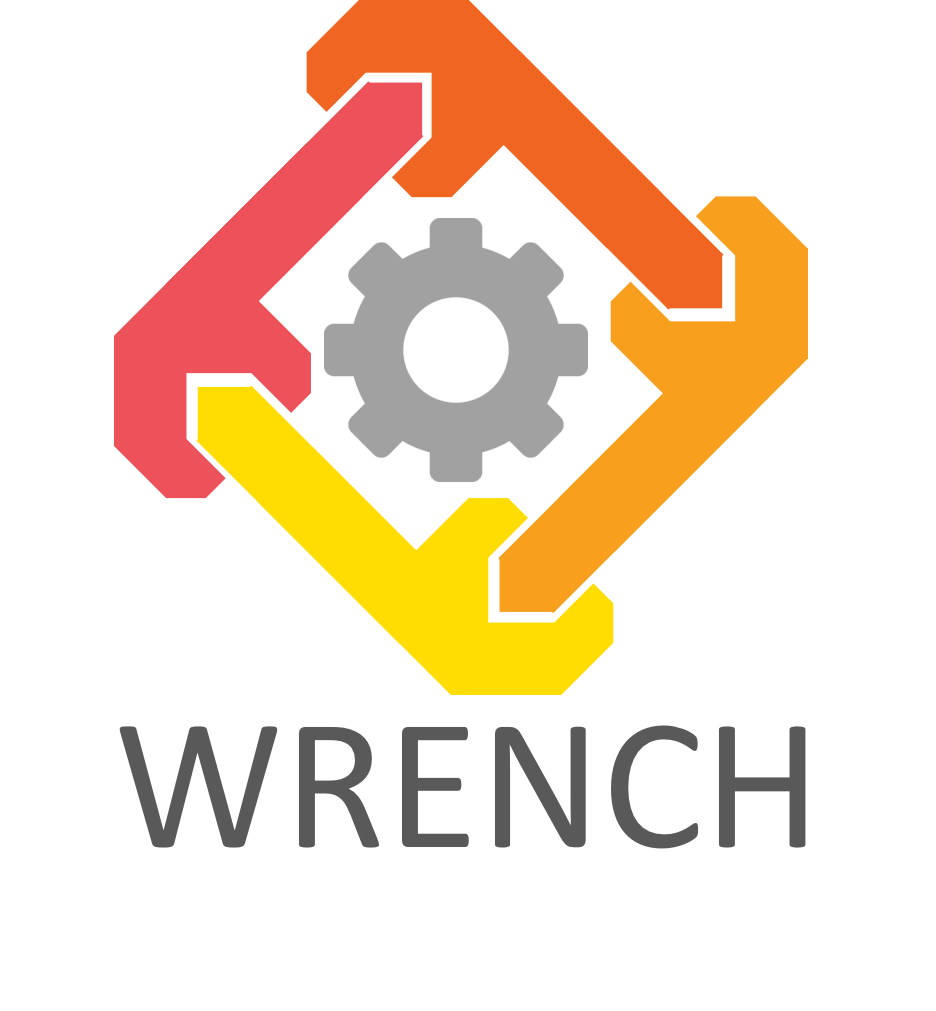}}
        \put(51,48){\small\textbf{DCSim}}
    \end{overpic}
    \caption{Job execution times per execution machine (worker node) vs. the hitrate in the real world (left) and the simulation (right). Plots show medians and 2.5- and 97.5-percentiles of all executed jobs on each execution machine.}
    \label{fig:validation}
\end{figure}

Once the simulator is calibrated it is validated against ground-truth data other than that used for performing the calibration.
When the simulated execution traces match those in the ground-truth data, we conclude that the simulation models, and their calibration, are valid.
A key performance metric, that is computed from real-world or simulated execution traces, is the job execution time, on each machine in the platform.
This metric depends highly on the fraction of input files that are initially in cache at the beginning of the execution (which we call the hitrate).
\cref{fig:validation} shows job execution times vs. hitrate, in the real-world and in simulation for a platform with three worker nodes. All worker nodes have the same architecture but two of them have 12 cores and the third has 24 cores. 

The calibrated simulation is able to reproduce the observed medians of the job execution times in the validation data without any adjustments to the simulation parameters.
However, as hitrate values increase, the simulated spread in the execution times is increasingly underestimated.
This is expected since the simulation does not capture all aspects of the real-world system.
In particular, the complex I/O access patterns for the hard disk drives that store cache data are not simulated.
These I/O accesses, due to seek times and random accesses, are non-deterministic.
The simulator, however, models I/O access times simply based on latencies and read/write bandwidth rates of the storage resources, which leads to a much lower variance of job execution times than in the real world.
For flash drive storage the seek times would be both smaller and more deterministic, which could be simulated with higher accuracy.

\subsection{Computational Complexity}
\label{sec:complexity}
\begin{figure}
    \centering
    \includegraphics[width=0.49\linewidth]{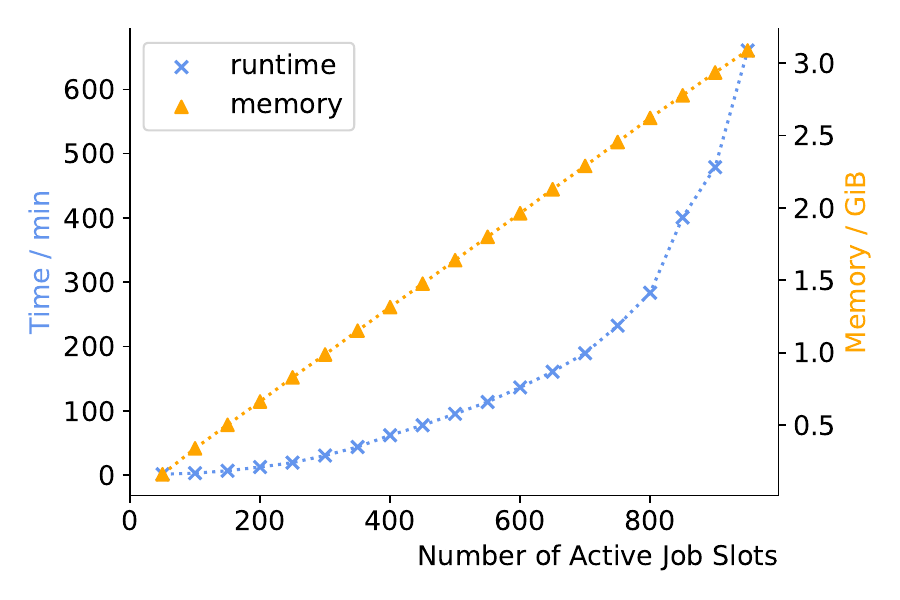}
    \begin{overpic}[width=0.49\linewidth]{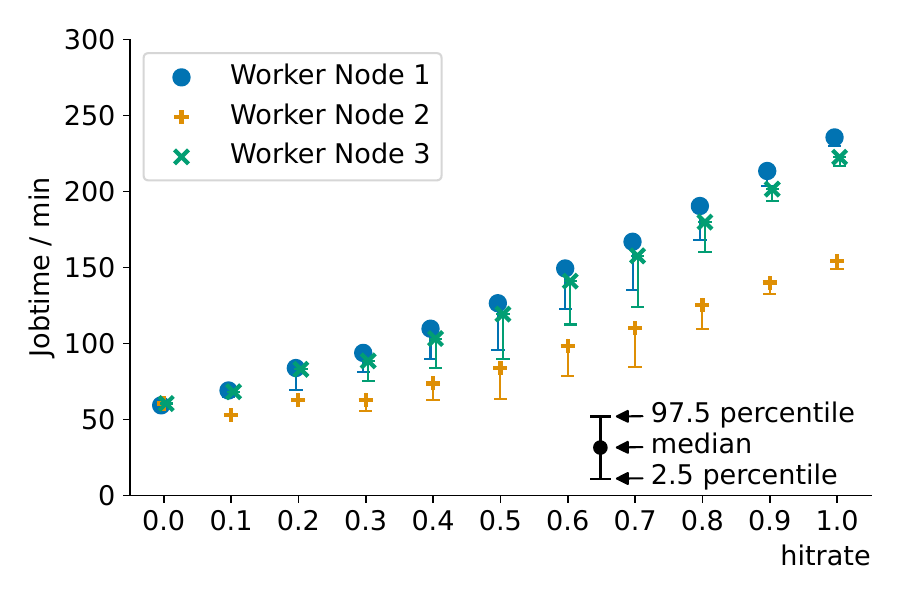}
            \put(50,54){\includegraphics[width=0.05\linewidth]{figures/Computing/SimGrid/simgrid_logo.pdf}}
            \put(60,50){\includegraphics[width=0.05\linewidth]{figures/Computing/SimGrid/wrench-logo-vertical.png}}
            \put(51,48){\small\textbf{DCSim}}
        \end{overpic}
    \caption{Simulation execution time and maximum memory consumption vs. the maximum number of active job slots, a proxy for the platform size (left); and simulated job execution times vs. hitrate  with mitigated time complexity (right).}
    \label{fig:scaling}
    \vspace{-1\baselineskip}
\end{figure}

To assess the space- and time-complexity of the simulator the maximum memory consumption and execution time of the simulation as functions of the size of the simulated platform are studied.
\cref{fig:scaling} shows the scaling of the simulation's memory consumption and execution time.
The simulation execution time is driven by the effective size of the platform, that is the maximum number of jobs that can execute concurrently in active \emph{job slots}.
A superlinear increase in execution time as platform size increases is observed.

Given the above, a mitigation procedure is needed to enable the simulation of large platforms under a time budget. 
By scaling the number of cores on each simulated machine by a constant factor the effective number of active job slots in the simulation of a large platform, and therefore the simulation execution time, can be decreased superlinearly.
To conserve the dynamics of the simulation the relative proportions of all other platform parameters are scaled by the same factor.
The mitigation procedure is applied on the simulation of the validation platform presented in \cref{sec:calandval}.
Comparing the simulation traces of the full simulation, visualized in \cref{fig:validation}, and mitigated simulation, visualized in \cref{fig:scaling}, no significant deviation can be observed.
Consequently, the mitigation procedure makes the simulation scalable while conserving the validity of the original simulation.


\section{Proof-of-Concept: Diskless Tier\,2}
\label{sec:poc}

To showcase the applicability of the presented simulator for modeling the performance of such architectures, we study two aspects of a future candidate design for a subset of the WLCG infrastructure proposed by the German HEP-computing community in~\cite{KET-Pos}.

\subsection{Simulated Workload and Platform}
\label{sec:workload}
The simulated workload characteristics for these studies are obtained from job monitoring data gathered by the CMS collaboration of jobs executed at the WLCG Tier\,1 site at KIT and Tier\,2 site at DESY in a time frame from February 24th to March 7th 2023.
From this data, and for each job, we extracted information about computational volume, amount of read and written data, consumed memory, and number of CPU cores used.

The jobs are classified as \textit{Analysis}, \textit{ReadoutSim}, \textit{Processing}, \textit{Merge}, and \textit{Other} jobs, which make up approximately 49.7\%, 39.5\%, 7.6\%, 3\% and 0.2\% of the workload, respectively.
\textit{Analysis} jobs contain all jobs submitted by individual users and therefore shows a large spread in the distributions of their job characteristics.
\textit{ReadoutSim} jobs process the simulation of the readout electronics of the CMS detector~\cite{CMS} and the digitization of the simulated signals.
\textit{Processing} jobs process the reconstruction from the raw digitized detector signals to analysis objects usable in analyses.
\textit{Merge} jobs combine multiple data files created in multiple jobs into single files simplifying subsequent data processing.
\textit{Other} jobs contain all generation, simulation, reconstruction, cleanup, and logging executions that did not occur often enough in the monitored time frame to result in a significant class.

From the individual characteristics of the jobs in each class we derived a one-dimensional binned probability distribution for each characteristic quantity.
Then, we sampled the characteristics of the simulated workload from these probability distributions while keeping the proportions relative to those observed in the monitoring data. We neglected the correlations between the probability distributions of the individual characteristics.

The simulated platform consists of two hypothetical sites approximately representing the real-world WLCG Tier\,1 and Tier\,2 sites at KIT and DESY.
Each site is modeled as its own local network of worker nodes and storage resources, and the two sites are interconnected via WAN.
The exact platform parameters are listed in \cref{tab:disklesstier2scenario}.

\begin{table} [hbtp]
    \centering
    \caption{Characteristics of the platform input into the simulation for the PoC study.}
    \label{tab:disklesstier2scenario}
    \begin{tabular}{c c c}
        \toprule
        Characteristic & Tier\,1 & Tier\,2\textquotesingle \\
        \midrule
        \midrule
        Compute & 42,000 cores & 20,000 cores \\
        Storage Bandwidth & \SI{80}{\giga\bit\per\second} & n/a\\
        Cache Bandwidth & n /a  & \SI{80}{\giga\bit\per\second} \\
        LAN Bandwidth & $2\times\SI{100}{\giga\bit\per\second}$ & $\SI{40}{\giga\bit\per\second}$ \\
        \midrule
        WAN Bandwidth & \multicolumn{2}{c}{\SI{100}{\giga\bit\per\second}}\\
        \bottomrule
    \end{tabular}
\end{table}

The two sites are considered to be isolated from the greater context of the WLCG, meaning that there is no additional load on the WAN caused by workloads running on other sites.

\subsection{Study 1: Grid Storage Replacement}
\label{sec:studyone}
The first simulation study evaluates the performance of a WLCG Tier\,2 site that is not a full-fledged storage site (which would require maintenance and a personnel budget) but instead serves only as a simple stand-alone cache with no availability or persistence guarantees (Tier\,2\textquotesingle\xspace). 
All input files in the simulation are served by the grid storage at the Tier\,1 site.
A defined fraction of input files are additionally staged at the cache at Tier\,2\textquotesingle\xspace at start of the simulation. This makes it possible to evaluate the performance of the workload execution dependent on the fraction of cached files (hitrate) with repeated simulations for multiple hitrate value between zero and one.
A hitrate~$=1$ corresponds to the nominal case of a Tier\,2 site following the WLCG definition.
A hitrate~$<1$ corresponds to the hypothetical scenario where the managed storage at the Tier\,2 is reduced to a cache with a performance of this cache parameterized via the hitrate.
At a hitrate~$=0$, there is no cache and all input files are served by the Tier\,1 storage.

\begin{figure}
    \centering
    \begin{subfigure}{0.49\linewidth}
        \includegraphics[width=0.99\linewidth]{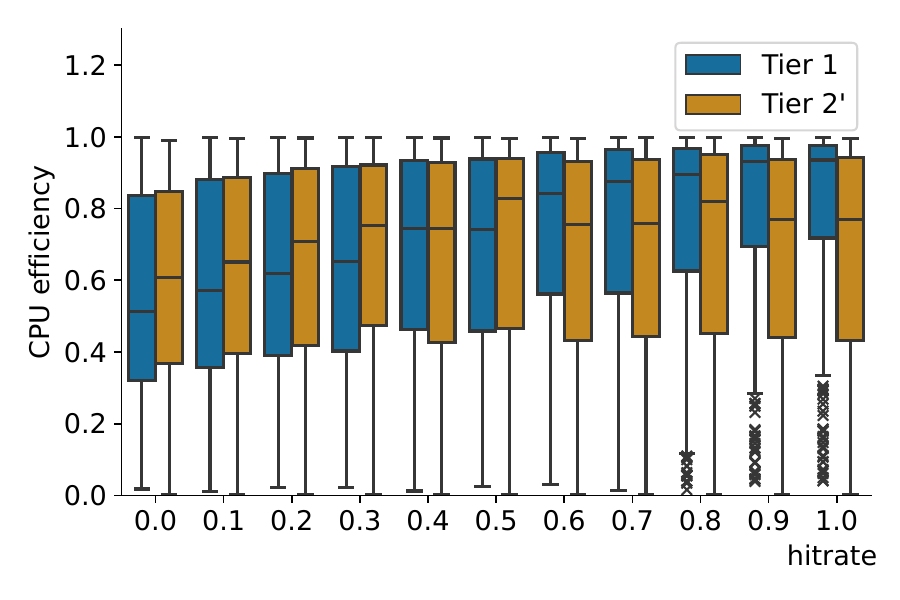}
        \caption{Study 1}
        \label{fig:disklesstier2result}
    \end{subfigure}
    \hfill
    \begin{subfigure}{0.49\linewidth}
        \includegraphics[width=0.99\linewidth]{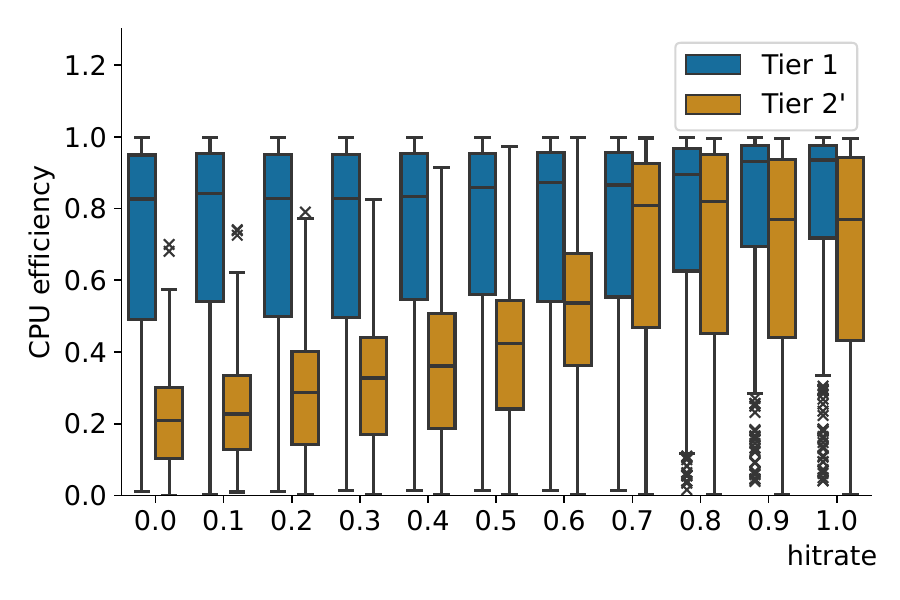}
        \caption{Study 2}
        \label{fig:hpcwithcacheresult}
    \end{subfigure}
    \vspace{-.75\baselineskip}
    \caption{Job CPU efficiency vs. the hitrate for the scenarios in \cref{sec:studyone} (left) and \cref{sec:studytwo} (right). Results 
    shown as standard box plots with the median, 25- and 75-quartiles, minimal and maximal single values within the interval spanned by the lower (upper) quartile minus (plus) 1.5 times the interquartile range, and outliers.}
    \label{fig:scenarioresults}
    \vspace{-.75\baselineskip}
\end{figure}

To estimate the performance of each site we measure the distribution of the fraction of time spent processing over the total job execution time (CPU efficiency) of the executed jobs.
The obtained distributions in simulation are visualized in \cref{fig:disklesstier2result}.

It can be observed in \cref{fig:disklesstier2result} that a replacement of the grid storage with a cache leads to a decrease in the performance of both sites.
For hitrates $<1$ the bandwidth to the grid storage at the Tier\,1 is shared by the jobs on both sites which increases the idle time of the CPUs due to the jobs waiting for data to process.
For hitrates $\gtrapprox0.5$ the performance of the Tier\,2\textquotesingle\xspace stabilizes due to the internal network throttling the execution.
The Tier\,2 load on the Tier\,1 storage still limits the performance of the Tier\,1.
To overcome the performance drop, an increase of the Tier\,1 storage bandwidth needs to be considered.

\subsection{Study 2: Cache Addition at High Latency Site}
\label{sec:studytwo}
In this study we explore the impact of replacing a dedicated Tier\,2 site by a hypothetical supercomputer.
Hence, the study presented in \cref{sec:studyone} is repeated but the WAN bandwidth between the sites is reduced by a factor of ten, which corresponds to a scenario in which the supercomputer does not have a high WAN bandwidth. 
This drastic reduction in the WAN bandwidth changes the dependence of the performance of both sites on the hitrate drastically.

The effect of the lower network bandwidth between the sites is that the execution time of jobs at Tier\,2\textquotesingle\xspace  is dominated by the time spent waiting for input data for hitrate $\lessapprox0.7$.
Above this value, job executions at Tier\,2\textquotesingle\xspace are once again throttled by the relatively low internal network bandwidth at that site. This leads to less load introduced by these jobs on the grid storage at Tier\,1.
As a consequence, job executions at Tier\,1 are only marginally affected, but the performance of job executions on Tier\,2\textquotesingle\xspace are significantly improved by high hitrates.

\section{Conclusions}

We have proposed using simulation to predict the performance of current and hypothetical architectural designs for distributed computing platforms such as the WLCG.
To this end we have implemented, calibrated, and validated an open-source simulator~\cite{Horzela2023} based on the SimGrid and WRENCH simulation frameworks.
The simulator's flexible configuration and its use of scalable simulation models make it possible to simulate and analyze the execution of a wide range of application workloads on large-scale computing platforms under an acceptable time budget.
These capabilities were demonstrated in two case studies, which 
involve systems with collectively more than 60,000 cores and \SI{10}{\peta\byte} of storage integrated into a distributed network, and applications transferring and processing petabytes of HEP data.

A future direction is the simulation of larger distributed computing systems up to the size of the WLCG and beyond.
This will require reducing the computational complexity of the underlying simulation models while preserving their accuracy.

\section*{Acknowledgements}
This work was supported by the German Federal Ministry of Education and Research (project FIDIUM 05H21VKRC2) and the Institute of Experimental Particle Physics, the Steinbuch Centre for Computing and GridKa at the Karlsruhe Institute of Technology, Germany; by National Science Foundation awards 2103489 and 2106059; and by Laboratory Directed Research and Development Strategic Hire funding No.\,11134 from Oak Ridge National Laboratory, provided by the Director, Office of Science, of the U.S. Department of Energy.

\bibliography{proceeding.bib}
%
%

\end{document}